\documentclass[manuscript]{aastex}
\usepackage{emulateapj5}
\usepackage{xcolor}

\usepackage{epsfig}
\usepackage{amsmath}
\usepackage{amssymb}
\usepackage{natbib}
\usepackage{graphicx}

\shorttitle{Observational implications of two families of compact stars}
\shortauthors{Bhattacharyya et al.}

\begin{document}

\title{Two coexisting families of compact stars: observational implications for millisecond pulsars}

\author{Sudip Bhattacharyya\altaffilmark{1}, Ignazio Bombaci\altaffilmark{2,3}, Domenico Logoteta\altaffilmark{3}, and Arun V. Thampan\altaffilmark{4,5}}

\altaffiltext{1}{Department of Astronomy and Astrophysics, Tata Institute of Fundamental Research, Mumbai 400005, India; sudip@tifr.res.in}
\altaffiltext{2}{Dipartimento di Fisica, Universit\`{a} di Pisa, Largo B. Pontecorvo, 3 I-56127 Pisa, Italy.}
\altaffiltext{3}{INFN, Sezione di Pisa, Largo B. Pontecorvo, 3 I-56127 Pisa, Italy.}
\altaffiltext{4}{Department of Physics, St. Joseph's College, 36 Lalbagh Road, Bangalore 560027, India.}
\altaffiltext{5}{Inter-University Centre for Astronomy and Astrophysics (IUCAA), India.}

\begin{abstract} 
It is usually thought that a single equation of state (EoS) model ``correctly''
represents cores of all compact stars. Here we emphasize that two families of compact stars,
viz., neutron stars and strange stars, can coexist in nature, and that neutron stars can get converted to strange stars
through the nucleation process of quark matter in the stellar center. 
From our fully general relativistic numerical computations of the structures of
fast-spinning compact stars, known as millisecond pulsars, we find that such a stellar conversion causes 
a simultaneous spin-up and decrease in gravitational mass of these stars.
This is a new type of millisecond pulsar evolution through a new mechanism, 
which gives rise to relatively lower mass compact stars with higher spin rates. This could have
implication for the observed mass and spin distributions of millisecond pulsars.
Such a stellar conversion can also rescue some massive, spin-supported millisecond pulsars from collapsing 
into black holes. Besides, we extend the concept of critical mass $M_{\rm cr}$ for the neutron star sequence    
\citep{Berezhianietal2003, Bombacietal2004} to the case of fast-spinning neutron stars, and point out that 
neutron star EoS models cannot be ruled out by the stellar mass measurement alone. 
Finally, we emphasize the additional complexity for constraining EoS models, for example, by
stellar radius measurements using X-ray observations, if two families of compact stars coexist.
\end{abstract}

\keywords{equation of state ---  pulsars: general --- pulsars: individual (PSR J1614-2230, PSR J0437-4715) --- stars: neutron --- stars: rotation}

\section{Introduction}\label{Introduction}

Millisecond pulsars (MSPs) are compact stars (CSs) with typical spin rates of several hundred Hz
\citep{BhattacharyaHeuvel1991}. They were first discovered 
in radio wavelengths, and thought to be spun up by accretion
in low-mass X-ray binaries \citep[LMXBs;][]{RadhakrishnanSrinivasan1982, Alparetal1982}.
Discovery of an accretion-powered X-ray MSP in an LMXB
system SAX J1808.4--3658 \citep{WijnandsKlis1998, ChakrabartyMorgan1998} gave credence to, and 
discoveries of transitional pulsars \citep{Archibaldetal2009, Papittoetal2013, 
deMartinoetal2013, Bassaetal2014} confirmed this idea. 
The database of radio and X-ray MSPs has steadily grown \citep[see, for 
example,][]{Lorimer2008, Watts2012, PatrunoWatts2012}. Recent years have seen precise 
measurement of masses of several MSPs (see \citet{Bhattacharyyaetal2017}
and references therein). Despite attempts to understand these observed mass
and spin rate distributions \citep[e.g., ][]{Ozeletal2012, Tauris2012, Kiziltanetal2013, Smedleyetal2014, 
Papittoetal2014, BhattacharyyaChakrabarty2017, Patrunoetal2017}, there still exist several questions:
do MSPs have two spin distributions and CSs have more than one birth mass distribution, 
why are observed MSP spin rates cut off around 730 Hz with a high confidence \citep{Chakrabartyetal2003, Patruno2010}, 
why are average spin rates of X-ray MSPs tentatively higher than those of radio MSPs \citep{Tauris2012}, and so on. 
Answering these questions is important to understand the CS formation,
spin-up and spin-down torques and accretion of these stars, stellar and binary evolution and
the stellar magnetic field. It is essential to identify all mechanisms of stellar evolution
to address these problems.

With densities in excess of the nuclear saturation density $\rho_{\rm sat} = 2.6 \times 10^{14}$~g/cm$^3$ in their interiors, 
what composes cores of CSs is another important question, which can be addressed by constraining (eliminating) theoretically 
proposed equation of state (EoS) models (see e.g., \citet{Bhattacharyya2010, Bhattacharyyaetal2017} for details on this).
This can be achieved by measuring at least three independent stellar parameter values
\citep[e.g.,][]{Bhattacharyya2010, Bhattacharyyaetal2017}. While the stellar mass and spin rate
have been precisely measured for a number of sources, a reliable measurement of a third parameter
has not yet been done. However, in the near future, the {\it NICER} satellite (which has been launched in June 2017) is expected to
measure the radius of at least one MSP with $\approx 5$\% accuracy using X-ray observations \citep{Gendreauetal2012}.
Besides, the moment of inertia of a CS in a double-pulsar system could be measured
in a few years \citep{Lyneetal2004}.

Considering the existence of two families of fast-spinning CSs, i.e. neutron stars (NSs) and strange stars 
(SSs), in this paper we illustrate the conversion process from the former to the latter for our example
EoS models. We also discuss the possible effects of such conversion on stellar mass and spin values
and on the procedure for constraining EoS models.

\section{EoS models and quark matter nucleation in neutron stars}\label{EoS}
In the present work we use the following EoS models.   
NS EoS models: the first NS EoS model (NS1) is the relativistic mean field EoS model TM1-2 \citep{providencia2013}, 
which includes nucleons and hyperons as baryonic degrees of freedom. 
The second NS EoS model (NS2) is the widely used APR EoS \citep{Akmaletal1998}, which includes 
only nucleons (as baryonic degrees of freedom) interacting via realistic two-body and three-body 
nuclear interactions.   
Note that the maximum gravitational masses for the non-spinning configuration are
$1.98 M_\odot$ and $2.19 M_\odot$ for NS1 and NS2 EoS models 
respectively, which are compatible with the measured maximum CS mass \citep[$2.01\pm0.04 M_\odot$; ][]{Antoniadisetal2013}.
In addition, neutron stars with the NS2 EoS model have smaller radii than stars with the NS1 EoS model. 
For example, for the gravitational mass $M_{\rm G} = 1.5 M_\odot$, the NS radius $R = 11.5$~km in the case of the NS2 EoS model,
while $R = 14.4$~km for the NS1 EoS model. Thus these two models are representative of NSs with 
``small'' and ``large''radii respectively.

SS EoS models: we consider an extended version of the MIT bag model EoS which includes 
perturbative corrections up to the second order in the strong structure constant \citep{Fra01, weis11}. 
We use two parameter sets for this model: the first with an effective bag constant 
$B_{\rm eff}^{1/4} = 140$~MeV and perturbative QCD corrections term parameter $a_4 = 0.8$ (SS1), 
and the second with $B_{\rm eff}^{1/4} = 125$~MeV, $a_4 = 0.5$ (SS2). 
We obtain the maximum masses of $2.04 M_\odot$ and $2.48 M_\odot$ for non-spinning configurations
of the SS1 model and the SS2 model respectively.
These two SS EoS models span the values of the parameters $B_{\rm eff}^{1/4}$ and $a_4$  
which are compaptible with the Bodmer--Witten hypothesis \citep{bod71,witt84} of absolute stability 
of strange quark matter, and give SS configurations which are compatible 
with the measured maximum CS mass \citep{Antoniadisetal2013} and the measured maximum CS spin
\citep[$\nu \approx 716$~Hz; ][]{Hessels2006}.

The mass-radius relations for non-spinning stars, for the NS1 and SS1 EoS models, are plotted in Fig. 1b.  
The conversion process of an NS to an SS
\citep[hereafter the NS$\rightarrow $SS conversion; ][]{BombaciDatta2000} and the existence of two families of compact stars
\citep[viz., NSs and SSs; ][]{Berezhianietal2003,Bombacietal2004,Bombacietal2008,Bombacietal2016,Dragoetal2016} are among the most prominent astrophysical consequences of having a first-order quark deconfinement phase transition in NS cores. 
This transition is possible when the Gibbs energy per baryon ($\mu$) versus pressure ($P$) curve 
for the hadronic-phase 
(HP, i.e., quarks confined within baryons and mesons; black curve in Fig. 1a)  
and that for the quark-phase (red curve in Fig.~1a) cross each other at the transition 
pressure $P_0$.     
For $P > P_0$ the HP is metastable and the QP will form via a nucleation process 
(in analogy to the fog or dew formation in supersaturated vapor, or ice formation in supercooled water). 

Small localized fluctuations in the state variables of the metastable HP 
will give rise to virtual drops of the stable QP. These fluctuations  are characterized 
by a time scale  $\nu_0^{-1} \sim 10^{-23}$ s. This time scale is set by the strong 
interactions (responsible for the deconfinement phase transition), and it is many orders of magnitude 
shorter than the typical time scale for the weak interactions.  
Quark flavor must therefore be conserved during the deconfinement transition 
\citep{ol94,Bombacietal2004,lug05}.    
We refer to this form of deconfined matter, in which the flavor content is equal to that of 
the $\beta$-stable hadronic system at the same pressure, as the Q*-phase 
\citep[Q$^*$P; ][]{Bombacietal2004}. 
Soon after a critical size drop of quark matter is formed, the weak interactions  
will have enough time to act, changing the quark flavor fraction of the deconfined droplet to lower 
its energy, and a droplet of $\beta$-stable quark matter is formed (hereafter the Q-phase).
This first seed of quark matter will trigger the NS$\rightarrow $SS conversion. 

Thus NSs with a central pressure $P_{\rm c} \ge P_{\rm cr} \ge P_{\rm 0}$ 
(central density $\rho_{\rm c} \ge \rho_{\rm cr} = \rho(P_{\rm cr})$) corresponding to a 
{\it short} (compared to typical pulsar ages) quark matter nucleation time $\tau(P_{c})$ 
will be converted to SSs and will thus populate the second branch of CSs. 
Following \citet{Berezhianietal2003,Bombacietal2004}, we take $\tau(P_{\rm cr}) = 1$~yr 
(which defines the {\it critical pressure} $P_{\rm cr}$ for stellar conversion)  and denote the 
corresponding NS configuration with the gravitational mass $M_{\rm cr} = M_{\rm G}^{\rm NS}(P_{\rm cr})$ as 
the {\it critical mass} configuration (black filled circle on the NS sequence in Fig.~1b).  

Since it is very unlikely to observe an NS with $M_{\rm G}^{\rm NS} > M_{\rm cr}$, $M_{\rm cr}$ plays the role of 
an effective maximum mass for the NS sequence \citep{Bombacietal2004}. 
The red filled circle on the SS1 curve (Fig.~1b) represents the SS, with gravitational mass $M_{\rm fin}$,  
formed by the conversion of the metastable NS with $M_{\rm cr}$. 
Following \citet{BombaciDatta2000}, we assume that during the stellar conversion process the total number 
of baryons in the star is conserved (i.e. we neglect any possible mass ejection), and so the stellar 
rest mass $M_{\rm 0}$ stays  constant. 
Therefore the total energy liberated in the stellar conversion is given by the difference between 
the gravitational mass of the initial NS ($M_{\rm in} \equiv M_{\rm cr}$) and that of the final SS $M_{\rm fin}$ 
configuration with the same rest mass (i.e., with $M_{\rm 0,cr} = M_{\rm 0,fin}\,$):  
$E_{\rm conv} = (M_{\rm in} - M_{\rm fin}) c^2 \sim 10^{53}$~ergs. 
This huge amount of released energy will cause a powerful neutrino burst \citep{BombaciDatta2000}.

We next consider the NS1 and SS2 EoS models for describing the HP and QP respectively.  
The corresponding Gibbs energies as a function of the pressure are plotted in Fig.~1c. 
As we can see in this case the two curves never cross each other, thus the quark deconfinement 
phase transition is not energetically possible and consequently it is not possible to populate the 
SS branch (Fig. 1d).

For the NS2 and SS1 EoS models and a 
quark matter surface tension $\sigma = 10$~MeV/fm$^2$, we find 
$M_{\rm cr}  = 2.16  M_\odot \simeq M^{\rm NS}_{\rm G,max} = 2.19 M_\odot$ and  
$M_{\rm fin} = 2.03  M_\odot \simeq M^{\rm SS}_{\rm G,max} = 2.04 M_\odot$
($M_{\rm G,max}$ implies the maximum non-spinning mass), 
thus there is a very small range of metastable NS configurations (with $2.16 \leq M_G/M_\odot \leq 2.19$) 
that can populate a tiny region (with $2.03 \leq M_{\rm G}/M_\odot \leq 2.04$) of the SS family.    
In addition we find that no stellar conversion occurs for $\sigma > 10$~MeV/fm$^2$
for the NS2+SS1 EoS model. 

For the NS2 and SS2 EoS models, we find $P_{\rm 0} > P_{\rm c}(M^{\rm NS}_{\rm G,max})$,
i.e., the transition pressure is larger that the central pressure of the NS maximum mass configuration.
Thus the NS family is stable for the NS2+SS2 EoS model. 

In summary, we find that NS2 stellar configurations are essentially stable with respect to the 
NS$\rightarrow $SS conversion when the strange quark matter phase is described by the 
extended MIT bag models of EoS \citep{Fra01, weis11} used in this work.

\section{Rapidly spinning stellar structure computation}\label{Computation}

In this work, we consider the effect of rapid spin on NS and SS configurations, and
on NS$\rightarrow $SS conversion.
Solving Einstein's field equations as detailed in \citet{Cooketal1994}, and also described
in \citet{Bombacietal2000, Bhattacharyyaetal2000, BhattacharyyaThampanBombaci2001},
we compute fully general relativistic configurations of fast spinning CSs,
and obtain stellar parameters like total angular momentum ($J$),
spin frequency ($\nu$), equatorial circumferential radius ($R$), gravitational mass ($M_{\rm G}$), 
rest mass ($M_{\rm 0}$), moment of inertia ($I$).

We obtain constant parameter sequences for $M_{\rm 0}$, $\nu$ and $\rho_{\rm c}$. 
Such sequences are useful for a number of reasons. 
For example, a CS spin-down due to electromagnetic (magnetic dipole)
radiation and/or gravitational radiation, keeps the rest mass conserved, and 
hence follows an $M_{\rm 0}$ sequence.
Such sequences are of two types (1) {\it normal} sequences which terminate at one end with
a non-spinning configuration; and (2) {\it supramassive} sequences which contain no
non-spinning configuration \citep{Cooketal1994}. 
The $M_{\rm 0}$ sequence corresponding to the maximum non-spinning mass distinguishes the 
{\it supramassive} region from the {\it normal} region. 
In a {\it supramassive} sequence, when the star becomes unstable to quasi-radial mode perturbations 
(for $\dfrac{\partial J}{\partial \rho_{\rm c}}\Big|_{M_{\rm 0}}$ $\ge 0$),
it collapses into a black hole \citep{Cooketal1994}. This does not happen in a {\it normal} sequence.
In addition to $M_{\rm 0}$ sequences, we also compute $\nu$ sequences
\citep{Bhattacharyyaetal2016}, because $\nu$ is known for all MSPs and hence such
a sequence can be useful to constrain EoS models. In addition, a $\rho_{\rm c}$ sequence
is useful to determine the critical mass $M_{\rm cr}(\nu)$ for spinning NSs
at which the NS$\rightarrow $SS conversion occurs (see Section~\ref{EoS}).

There are four limits that define the stable stellar parameter space for a given EoS model
\citep{Cooketal1994}. These are
(1) the non-spinning limit implying $\nu \rightarrow 0$ and $J \rightarrow 0$;
(2) the mass-shed limit, at which matter cannot remain bound to the stellar surface
due to a very fast spin rate;
(3) the instability limit corresponding to the quasi-radial mode perturbations mentioned above; and
(4) the low-mass limit, below which a CS does not form.
We use these limits to ensure that our computed parameter values before and after the
NS$\rightarrow $SS conversion are within the stable stellar parameter space.

\section{Results and implications}\label{Results}

\subsection{Structures and limiting masses for the neutron star EoS}\label{NS}

In the rest of the paper, we use EoS models NS1 and SS1 for the purpose of demonstration,
because, among the four conversion options considered in this paper, only the NS1 to SS1
conversion is realistically possible (see Section~\ref{EoS}).
We begin with the NS EoS model structures (Section~\ref{Computation}) 
for which the $M_{\rm G}$ versus $\rho_{\rm c}$ curves for the non-spinning
limit and the mass-shed limit are shown in Figure~\ref{fig2}. 
The stable configurations in the parameter space between these two curves can support masses up
to $1.98 M_{\odot}$ (non-spinning) and $2.37 M_{\odot}$ (mass-shed limit).
The NS$\rightarrow $SS conversion, however, will modify this stable parameter space as described below.

A quark deconfinement phase transition from the
NS1 EoS model to the SS1 EoS model is allowed (Section~\ref{EoS} and Figure~\ref{fig1}a).
For these EoS models and a quark matter surface tension $\sigma = 10$~MeV/fm$^2$, 
the critical pressure, given a quark matter nucleation time $\tau(P_{\rm cr}) = 1$~yr,  
corresponds to a critical density $\rho_{\rm cr} = 6.8\times10^{14}$~g cm$^{-3}$. 
Regardless of the spin rate, this critical density relates to bulk matter properties in both phases
and to interface properties ($\sigma$) such that all spinning NS models with 
$\rho_{\rm c} > \rho_{\rm c, cr}$ (the critical central density $\rho_{\rm c, cr} = 
\rho_{\rm cr}$) will undergo NS$\rightarrow $SS conversion.
The blue vertical line of Figure~\ref{fig2} represents the value of the critical mass
configuration for spinning NSs:  $M_{\rm cr}(\nu) \equiv M_{\rm G}^{\rm NS}(\rho_{\rm c, cr}, \nu)$  
and provides a new stability limit, in addition to the four limits mentioned in Section~\ref{Computation}.
The corresponding values for the critical mass $M_{\rm cr}(\nu)$ are $1.61 M_{\odot}$ and $2.03 M_{\odot}$
for the non-spinning limit and the mass-shed limit ($925$~Hz) respectively.  
In this mass and $\nu$ range, we use 19 points for the computation of 
NS$\rightarrow $SS conversion (Figure~\ref{fig2}).  
Among these, the red points are in the {\it normal} region and the green point is in the {\it supramassive} region.

\subsection{Conversion of a neutron star into a strange star}\label{SS} 

Focussing on observationally relevant parameters (see Section~\ref{Introduction}),  
we investigate the conversion of fast-spinning NSs to fast-spinning SSs by studying how 
$M_{\rm G}$, $R$, $I$ and $\nu$ change.
The NS$\rightarrow $SS conversion happens when the NS attains a central density equal to 
$\rho_{\rm c, cr}$ ($= 6.8\times10^{14}$~g cm$^{-3}$ in our case; see Section~\ref{NS}) 
via evolution (e.g., by mass accretion or electromagnetic spin-down). 
We assume that the conversion preserves rest mass $M_0$ and angular momentum $J$.
The pre-conversion configuration (corresponding to $M_{\rm cr}$) for the NS1 model is computed 
for $\rho_{\rm c} = 6.8\times10^{14}$~g cm$^{-3}$ and a given $\nu$ value,
and the corresponding $M_{\rm 0}$ and $J$ values are calculated (see Section~\ref{Computation}).
The post-conversion parameter values are found by computing the stable stellar structure for the 
SS1 EoS model for the above mentioned $M_{\rm 0}$ and $J$ values.
The conversion results are shown in Figure~\ref{fig3}.

We find that as a result of the conversion, $M_{\rm G}$, $R$ and $I$ always decrease,
while $\nu$ increases (except for $\nu = 0$).
Therefore, an NS$\rightarrow $SS conversion spins up the star,
while making it less massive. Typically, $\nu$
increases by $\approx 21-35$\%, while $M_{\rm G}$ decreases by $\approx 7-9$\%,
and $R$ and $I$ decrease by $\approx 20-24$\% and $\approx 17-26$\% respectively.

The green point to a gray point transition in Figure~\ref{fig3} displays a {\it supramassive} NS regime 
to {\it normal} SS regime conversion.
Note that in the standard single stellar family scenario, in the {\it supramassive} regime, 
a CS's expected fate is to collapse into a black hole by reaching the instability
limit through spin-down (Section~\ref{Computation}). 
Therefore, our result shows a new mechanism to rescue a CS from this fate.

We have made an assumption that during the stellar conversion process the total number of baryons 
    in the star is conserved. If there is some ejection of mass, the decrease of the stellar gravitational mass 
    due to the conversion process will be larger than the one calculated assuming $M_0 = {\rm const}$. 
    $J$ may change in this case too. We do not consider this possibility in the present work.     
However, even assuming $M_0 = {\rm const}$ during the conversion process, the released neutrinos 
may take away significant amount of angular momentum. 
From \citet{Epstein1978} and the computed fractional loss of $M_{\rm G}$, we estimate that at most 
20\% of the stellar angular momentum could be lost due to the neutrino emission.  
In Figure~\ref{fig3}, we show two final configurations for 20\% loss of $J$, 
but with $M_{\rm 0}$ conserved, as blue squares.  
Our qualitative conclusions remain the same even for this maximum angular momentum loss. 

These findings have important implications for MSPs.
As described in Section~\ref{Introduction}, for understanding the observed mass and spin distributions 
of MSPs, it is important to identify all mechanisms that cause these two parameters of CSs to evolve.
In Figure~\ref{fig4}, we give cartoons of main mechanisms.
{\it Panel a} of Figure~\ref{fig4} illustrates the
increase of $\nu$ with increasing $M_{\rm G}$, due to accretion 
during most of the LMXB phase. This is the primary mechanism through which
a CS spins up substantially and becomes an MSP \citep{BhattacharyaHeuvel1991}.
{\it Panel b} shows another type of evolution where $\nu$ decreases with increasing $M_{\rm G}$: 
happening at the end of the LMXB phase, termed Roche-lobe decoupling phase (RLDP),
when the accretion rate decreases dramatically as the companion star
slowly decouples from its Roche-lobe. In this phase, there is an overall spin down
due to a net negative torque in the so-called ``propeller'' regime \citep{Tauris2012},
although some mass can intermittently fall on the stellar surface \citep[e.g.,][]{DAngeloetal2010}.
A third type of evolution involves a decrease of $\nu$ with decreasing $M_{\rm G}$
({\it panel c}). This happens when the MSP spins down
by emission of electromagnetic radiation and/or gravitational radiation, keeping its rest mass
conserved and losing gravitational mass (see Section~\ref{Computation}).
In this paper, we report a ``complementary'' fourth type of evolution, representing increasing $\nu$ with 
decreasing $M_{\rm G}$ ({\it panel d}). Such an evolution can happen
when a fast-spinning neutron star becomes a fast-spinning strange star through
the conversion process described herein. 

In principle, there is an alternate way to achieve this fourth type of evolution:
when the CS evolves along a rest mass sequence very close to the instability limit in the {\it supramassive} region, 
its $\nu$ can increase with the decrease of $M_{\rm G}$ \citep[and with the decrease of $J$;][]{Cooketal1994}. 
But in reality, observed MSPs have a small probability to populate such a small parameter space 
with high masses and spin rates.
Therefore, our numerical computation in this paper provides not only a new mechanism of
significant spin-up, but also an accompanying unique gravitational mass decrease.
While within the scope of the calculations presented here, this new way of evolution
could significantly affect the mass and spin distributions of MSPs, the magnitude of this effect
will be better understood from a future detailed population 
study including this NS$\rightarrow $SS conversion mechanism.

\subsection{Constraining a pair of EoS models}\label{constraint}

We now discuss some implications of the existence of two families of CSs
on constraining EoS models using observational data. 

First we consider EoS model constraints from the measurement of $M_{\rm G}$ alone.  
A CS PSR J0348+0432 has high precisely measured $M_{\rm G} = 2.01\pm0.04 M_\odot$
\citep{Antoniadisetal2013}, which is thought to eliminate all EoS models that cannot support a mass 
$\sim 2 M_\odot$. But, if there exist in nature two families of CSs,
then it is sufficient that only the SS EoS model supports the highest measured mass.  
It is not necessary for the NS EoS model, the EoS before the 
the NS$\rightarrow $SS conversion, to support all measured masses.
Therefore, an NS EoS model cannot be ruled out by mass measurement alone.

For spinning NSs, we define the effective maximum mass 
to be that critical mass $M_{\rm cr}(\nu)$ for which  NS$\rightarrow $SS conversion takes place.
This mass value is smaller than the traditional maximum mass (see Figure~\ref{fig2}).
As mentioned before, $M_{\rm cr}(\nu)$ is related to the bulk properties of matter in both phases
(i.e., confined and deconfined) and to the interface properties.   
 
Finally, we explore a way to constrain EoS models, when two families of CSs exist.
For the case of a single stellar family, one could check the validity of the EoS model by
computing the $M_{\rm G}-R$ curve for the known $\nu$ value of
an MSP and then comparing with its measured $M_{\rm G}$ and $R$ values \citep{Bhattacharyyaetal2017}.
In a similar manner, for two families of CSs, the $M_{\rm G}-R$ curves for both NSs
and SSs are computed.  Figure~\ref{fig5} shows two such pairs of curves for the MSPs PSR J0437--4715 ($\nu = 173.6$~Hz)
and PSR J1614--2230 ($\nu = 317.5$~Hz). These CSs with measured $M_{\rm G}$ 
are possibly the most promising MSPs for radius measurements with
the {\it NICER} satellite \citep{Gendreauetal2012, Miller2016}. 
In order to meaningfully compare theoretical results with observations, it is essential to also compute 
the critical mass $M_{\rm cr}(\nu)$, or the effective maximum NS mass, at the measured $\nu$ value of both MSPs.
The effective maximum masses for our NS1 EoS models are 
$1.62 M_\odot$ and $1.64 M_\odot$ for PSR J0437--4715 and PSR J1614--2230 respectively.  
Similarly, the computed post-conversion minimum masses ($M_{\rm fin}(\nu)$) 
for the SS branch for our EoS models are $1.48 M_\odot$
and $1.49 M_\odot$ for PSR J0437--4715 and PSR J1614--2230 respectively (Figure~\ref{fig5}).
Therefore, higher mass MSPs can generally be used to constrain strange star EoS models,
while lower mass MSPs are ideal to probe neutron star EoS models.
MSPs in the overlapping mass ranges of the two stellar families 
(e.g., $\sim 1.48-1.64 M_\odot$ in Figure~\ref{fig5}) can be used to constrain the critical mass values
and thus the region of dense matter EoS where the deconfinement phase transition takes place.

\section{Conclusion}\label{Conclusion}

In this paper, we have explored some astrophysical consequences of the occurrence of a first-order quark 
deconfinement phase transition in CSs by quark matter nucleation, causing two coexisting CS families. 
Our fully general relativistic numerical computations for fast-spinning stars
show that the NS$\rightarrow $SS conversion causes a spin-up and a gravitational mass decrease.
This is, a new type of MSP evolution through a new mechanism, generating relatively lower mass stars
with higher spin rates, and also high mass SS products with significantly smaller radii.
This could have considerable contribution to the two possible spin populations of MSPs
and birth mass populations of CSs, as indicated by \cite{Patrunoetal2017, Ozeletal2012}.
The pre-conversion stellar mass plays the role of an effective maximum mass for the NS sequence.
Thus, a neutron star EoS model cannot be ruled out by the stellar mass measurement alone.

\begin{figure}[h]
\centering
\includegraphics*[width=10.0cm,angle=270]{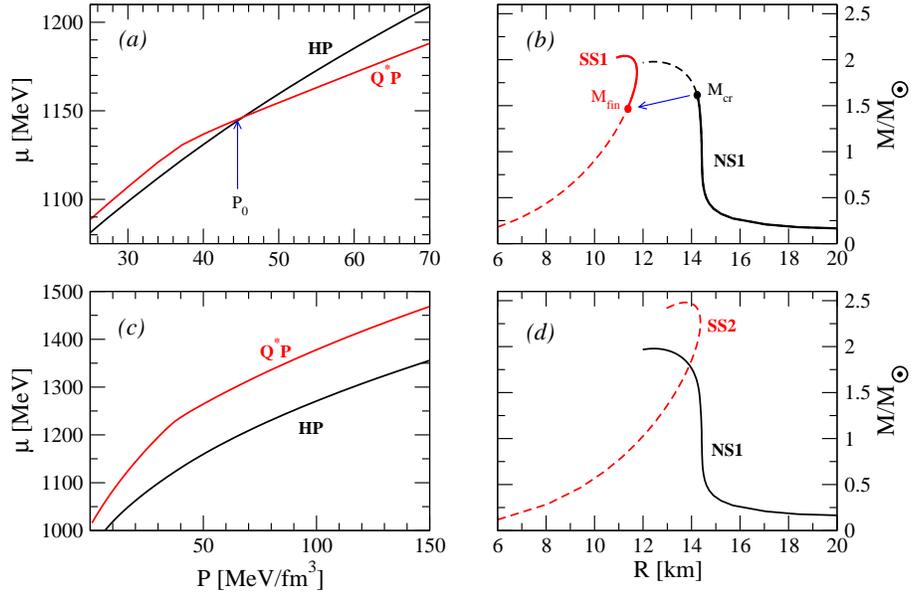} 
\caption{Left panels: Gibbs energy per baryon ($\mu$) versus pressure for the hadronic-phase (HP) 
         and the Q*-phase (Q$^*$P), i.e. the deconfined quark phase in which the flavor content is 
         equal to that of the $\beta$-stable hadronic system at the same pressure 
         (see \citet{Bombacietal2004} for more details).  
         Right panels ({\it b}): mass-radius relation for non-spinning neutron stars (NS) and strange stars (SS). 
         The two upper panels are relative to the SS1 EoS, whereas the two lower panels to the SS2 EoS. 
         The NS1 EoS model \citep{providencia2013} has been used in both cases.  
         The NS$\rightarrow $SS conversion is represented by the filled circles connected by an arrow
	 (Section~\ref{EoS}). 
\label{fig1}}
\end{figure}

\begin{figure}[h]
\centering
\includegraphics*[width=8.0cm]{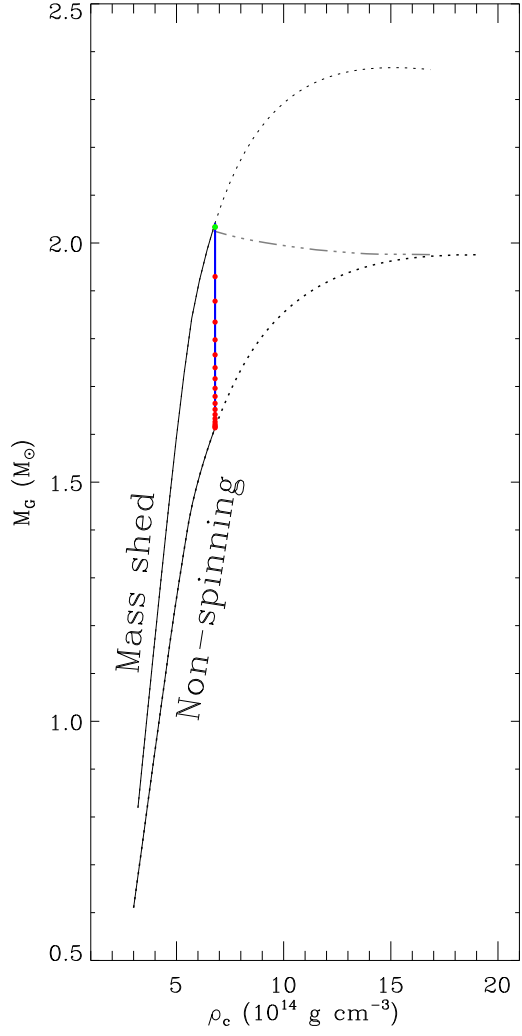}
\caption{$M_{\rm G}$ versus $\rho_{\rm c}$ curves 
	for non-spinning and mass-shed limits for neutron stars (NS1 EoS model). 
	The dash-triple-dot curve, the rest mass sequence 
	corresponding to the maximum non-spinning mass, distinguishes the {\it supramassive}
	region above from the {\it normal} region below. The blue vertical 
	line corresponds to the critical central density 
	$\rho_{\rm c, cr} = 6.8\times10^{14}$~g cm$^{-3}$, which 
	is the NS central density for NS$\rightarrow$SS conversion.     
	The red ({\it normal}) and green ({\it supramassive}) points on this line 
	show the initial NS mass values for transitions computed in this paper.  
	For our pair of EoS models (NS1 and SS1), the parameter space to the right of the blue line, including
	the dotted portions of the non-spinning and mass-shed curves, 
	cannot be attained by a neutron star (Section~\ref{NS}).
\label{fig2}}
\end{figure}

\begin{figure}[h]
\centering
\includegraphics*[width=9.0cm]{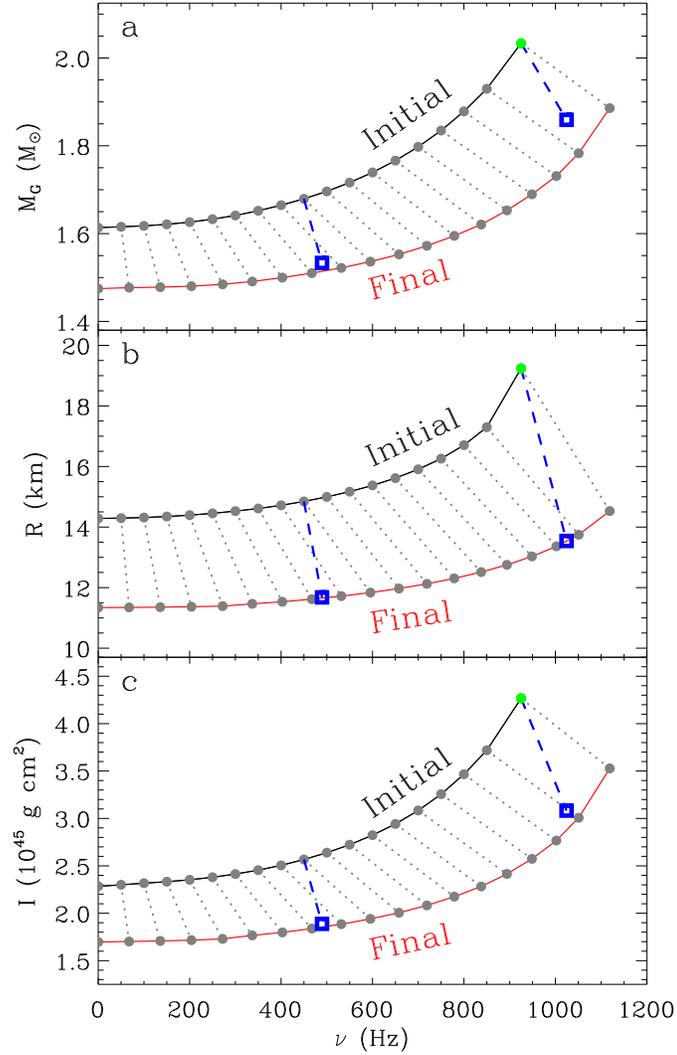}
\caption{NS$\rightarrow $SS conversion (for NS1 to SS1). 
	Each pair of initial and final points (filled circles) are connected with a dotted gray line.
	The gray filled circles are for the {\it normal} region, while the green filled circle is
	for the {\it supramassive} region. The filled circles for the NSs (marked by ``Initial")
	are same as those shown in Figure~\ref{fig2}. 
	While for most NS$\rightarrow $SS conversions, the stellar angular momentum 
	is assumed to be conserved, for two cases (shown with blue squares and dashed lines) 
	20\% angular momentum loss due to neutrino emission is considered. {\it Panels a}, {\it b}
	and {\it c} are for gravitational mass ($M_{\rm G}$) versus spin frequency ($\nu$),
	equatorial radius ($R$) versus $\nu$ and moment of inertia ($I$) versus $\nu$
	respectively. This figure shows that $M_{\rm G}$, $R$ and $I$ decrease and $\nu$
	increases, even for an angular momentum loss, for NS$\rightarrow $SS conversion
	(Section~\ref{SS}).
\label{fig3}}
\end{figure}

\begin{figure}[h]
\centering
\includegraphics*[width=9.0cm]{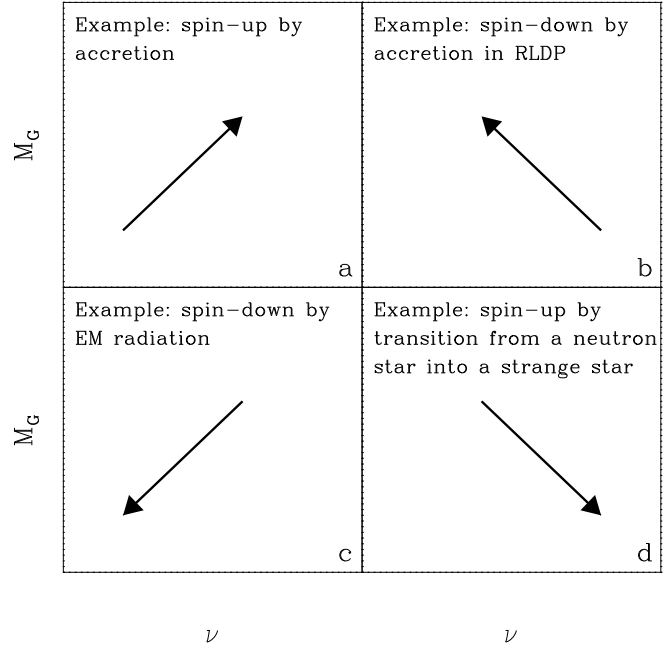}
\caption{An illustration showing all four possibilities of gravitational mass ($M_{\rm G}$) 
	versus spin frequency ($\nu$) evolution of MSPs in four panels. 
	{\it Panel a}: both $M_{\rm G}$ and $\nu$ increase, for example, via accretion in an LMXB.
	{\it Panel b}: $M_{\rm G}$ increases but $\nu$ decreases, for example, in the Roche
	lobe decoupling phase (RLDP) towards the end of the lifetime of an LMXB.
	{\it Panel c}: both $M_{\rm G}$ and $\nu$ decrease, for example, via electromagnetic
	(magnetic dipole) radiation.
	{\it Panel d}: $M_{\rm G}$ decreases but $\nu$ increases, for example, via 
	NS$\rightarrow $SS conversion (Section~\ref{SS}).
\label{fig4}}
\end{figure}

\begin{figure}[h]
\centering
\includegraphics*[width=9.0cm]{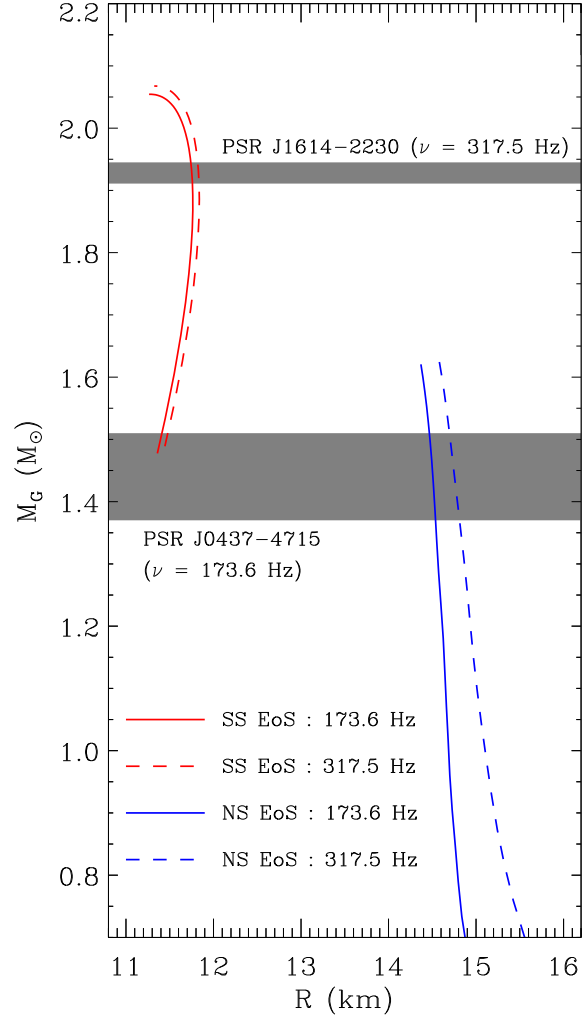}
\caption{$M_{\rm G}$ versus $R$ curves for NS1 (blue) and SS1 (red) EoS configurations.  
        Two curves, one for $\nu = 173.6$~Hz and another for $\nu = 317.5$~Hz, are given for each EoS model.  
  	Note that the conversion to the SS1 configurations (red) happen from NS1 configurations of 
        lower $\nu$ values (not shown in this figure), and {\it not from the blue curves}. 
        This is because spin-up happens due to the stellar conversion (see Figure~\ref{fig3}). 
	Two gray horizontal bands show the measured $M_{\rm G}$ ranges \citep{Reardonetal2016, Fonsecaetal2016} of two MSPs: 
	PSR J0437--4715 with $\nu = 173.6$~Hz and PSR J1614--2230 with $\nu = 317.5$~Hz.
	This figure suggests that, due to two disconnected segments of the $M_{\rm G}-R$ curve 
	for a known $\nu$ value, $M_{\rm G}$ and $R$ measurements of several 
	sources are required to probe the core matter of CSs
	(Section~\ref{constraint}).
\label{fig5}}
\end{figure}

Regarding constraints on EoS models, the advanced and innovative instruments on board the {\it NICER} satellite
will be able to determine MSP radii with high precision ({\it i.e.} better than
$\approx 1$~km uncertainty). Moreover, the masses of a number of MSPs are already known (\citet{Bhattacharyyaetal2017} and
references therein). These could potentially be useful to reasonably constrain 
EoS models of a single stellar family. For example, \citet{OzelPsaltis2009} suggested that parameter
measurements of three neutron stars could effectively distinguish some proposed EoS models.
But the two disconnected segments of the mass-radius curves, mostly in two different mass and radius ranges,
for two coexisting stellar families imply the necessity of mass and radius measurements of more number of MSPs. 
This is because, for two coexisting families,
one needs to constrain two separate EoS models, as well as to identify the NS configuration for conversion.
But even if a detailed characterization of two plausible mass-radius curves cannot be done in the near future,
only two CSs with slightly different measured mass values but with very different measured radius values
could be strongly suggestive of NS$\rightarrow $SS conversions discussed in this paper, unless these 
measured masses are close to the observed maximum mass (where the radius can significantly change
with a small change in mass; see Fig.~\ref{fig1}). 

\acknowledgments
We thank an anonymous referee for constructive suggestions.
AVT acknowledges funding from SERB-DST (File No.: EMR/2016/002033). 
This work has been partially supported by ``NewCompstar'', COST Action MP1304.


\clearpage

\clearpage

\clearpage

\clearpage

\clearpage

\end{document}